\documentclass{aa}
\usepackage{graphicx}
\usepackage{txfonts}
\usepackage{natbib}
\usepackage{multirow}

\begin{document}

  \title{Pulse to pulse flux density modulation from pulsars at 8.35 GHz}
  \titlerunning{Pulse to pulse flux density modulation from pulsars at 8.35 GHz}
  \authorrunning{O. Maron et at.}
  \author{O. Maron\inst{1}, M. Serylak \inst{2,3}, J. Kijak \inst{1}, K. Krzeszowski \inst{1}, D. Mitra \inst{4} \and A. Jessner \inst{5}}

  \offprints{O. Maron,\\ \email{olaf@astro.ia.uz.zgora.pl}}

  \institute{Kepler Institute of Astronomy, University of Zielona G\'ora, ul. Lubuska 2, 65-265 Zielona G\'ora, Poland
  \and
             Station de Radioastronomie de Nan\c{c}ay, Observatoire de Paris, CNRS/INSU, 18330 Nan\c{c}ay, France
  \and
             Laboratoire de Physique et Chimie de l'Environnement et de l'Espace, LPC2E UMR 7328 CNRS, 45071 Orl\'{e}ans Cedex 02, France
  \and
             National Centre for Radio Astrophysics, P.O. Bag 3, Pune University Campus, Pune, 411 007, India
  \and
             Max-Planck-Institut f\"ur Radioastronomie, Auf dem H\"ugel 69, D-53121 Bonn, Germany
            }
  \date{Received ; accepted }

  \abstract
  {}
  {To investigate the flux density modulation from pulsars and the existence of specific behaviour of modulation index versus frequency.}
  {Several pulsars have been observed with the Effelsberg radio telescope at 8.35 GHz. Their flux density time series have been corrected for interstellar scintillation effects.}
  {We present the measurement of modulation indices for 8 pulsars. We confirm the presence of a critical frequency at $\sim$1 GHz for these pulsars (including 3 new ones from this study). We derived intrinsic modulation indices for the resulting flux density time series. Our data analysis revealed strong single pulses detected from 5 pulsars.}
  {}

  \keywords{pulsars: general - pulsars: individual:B1133+16}

  \maketitle

  \section{Introduction}

    A typical pulsar spectrum is steep compared to spectra of other non-thermal radio sources and can be described by a power law $(S\propto \nu^{\alpha})$ in the radio frequency range. The average spectral index of pulsars is $-1.8$ \citep{mkk+00} which indicates that pulsars are weak radio sources when observed at high frequencies. According to radius-to-frequency-mapping \citep{cor78, kg03}, high frequency radiation originates close to a pulsar surface. Therefore, single pulse observations of pulsars at short wavelengths can provide valuable information for studies of pulsar radio emission close to the neutron star surface.

    Pulsar radiation undergoes various propagation effects as it passes through interstellar medium. Depending on the strength of the turbulence and homogeneity of the scattering material these effects cause modulation of phase of the propagating pulsar signal. This modulation manifests itself by variations is signal intensity over large range of observation time scales and bandwidths. These modulations are correlated over a characteristic scintillation bandwidth which is exponentially proportional to observing frequency, ($\Delta f \propto f^4$) as presented by \citet{sch68}. Scintillation effects are generally more pronounced at frequencies below 1 GHz. However, observing pulsars at high frequencies can also serve as a probe of properties of the interstellar medium as shown by \citet{bkg+00} and \citet{lkgk11}. These authors used high-frequency observations of pulsars to study the structure and dynamics of the local interstellar medium and direct determination of the interstellar electron density spectrum.

    The pulse-energy distribution can be used to investigate different possible models of pulsars emission physics. In a series of papers, Cairns and collaborators (\citealt{cjd01, cdr+03, cjd03, cjd04}) investigated different possible models of the emission physics using observations of three pulsars. They have shown that the pulse-energy distributions appear to be log-normal and are representative of the normal pulsar radio emission, can be fit with a single emission model, or convolution of Gaussian-log-normal or double log-normal models. The presence of approximately power-law distributions was ascribed to be due to the emission of ``giant pulses'', i.~e. pulses which have integrated flux density greater than 10 times the integrated flux density of the average profile. These pulses are believed to be associated with the high-energy emission in the outer magnetosphere (\citealt{jr02}, \citealt{cai04}).

    Also, measuring modulation index can be used to study the physical processes that create pulsar radio emission. \citet{jg03} used pulse-to-pulse modulation index to test different radio emission models. They used four ``complexity parameters'', which represent: sparking gap model \citep{gs00}, continuous current outflow instabilities model \citep{as79, ha01}, surface magneto-hydrodynamic wave instabilities model \citep{lou01} and outer magnetospheric instabilities model \citep{jg03}. The anti-correlation of those parameters with measured pulse-to-pulse modulation indices of a sample of 12 pulsars lead them to dismiss the surface magneto-hydrodynamic wave instabilities model. The results from survey of pulsars at 21 and 92 cm made by \citet{wes06, wse07} have shown that the modulation index increases at lower frequencies however; despite large sample of pulsars, none of the models could be favoured over others.

    Analysis of pulse-to-pulse modulation index shows, that the degree of modulation depends on the observing frequency, with an apparent minimum around 1 GHz for the majority of analysed pulsars \citep{bsw80}. In this paper we argue that this phenomenon is not intrinsic to pulsar emission mechanism, but it depends on sensitivity of the observations and presence of so called pseudo-nulls. The outline of this paper is as follows: we summarise our observations and data reduction done for a set of 12 pulsars, followed by our flux density measurements and corrections for the effect of interstellar scintillation. Then, we derive intrinsic modulation indices. Next, we compare the pulse-energy distributions from our observations with those made at other frequencies. We also report the detection of numerous strong single pulses from PSR B1133+16 and from 4 other sources albeit less abundant. Finally, we discuss our results in the last section.

  \section{Observations and data reduction}

    We carried out our observations with the 100-meter radio telescope of Max-Planck-Institut f\"ur Radioastronomie at Effelsberg in 2002 (March and December) and 2004 (April and June). The observations were made using the cooled secondary focus receiver with HEMT amplifiers at the observing frequency of 8350 MHz and bandwidth of 1100 MHz. The receiver provided LHC and RHC signals that were digitised and independently sampled into 1024 pulse phase bins synchronously folded with the topocentric pulse period P, using
the Effelsberg Pulsar Observation System  (EPOS, \citealt{jes96}) which recorded the calibrated data in single pulse mode and stored for future off-line analysis. The aim of our observations was to detect single pulses from pulsars at this frequency. 

    In order to calibrate the received pulsar signal at the Effelsberg radio telescope, a noise diode is installed in every receiver. This diode is switched on synchronously with every pulse period. The signal output of the noise diode is then compared with the energy received from the pulsar. The energy from the noise diode is calibrated by comparing its output to the flux density of a known continuum source. This pointing procedure was performed on well known flux calibrators during our observations. For the 2002 observations we have used the following flux calibrators: 3C295, 3C147, 3C279, while for 2004 observations: 3C273, 3C196, NGC7027 and 3C286. In order to estimate the mean flux density $S_{\mathrm{mean}}$ (which represents the total on-pulse energy {\it E}, averaged over an entire pulse period {\it P}, i.e. {$S_{\mathrm{mean}}=E/P$}), we have used the calibration procedure which is described in detail by \citet{kra95}. The pulsar signal was sampled by dividing the pulsar period into 1024 bins of the entire pulse window. For observations of PSR B1133+16 made in April and June 2004 the resolution was increased to 60 ${\mathrm{\mu s}}$ by using the part of the pulsar period enclosing the pulse profile and dividing it into 1024 bins. Finally, the flux-calibrated time series were produced for further analysis. Table~\ref{table1} summarises the parameters of the observed pulsars.

    \setlength{\tabcolsep}{5pt}
    \begin{table*}\footnotesize
      \begin{minipage}{\textwidth}
        \centering
        \caption[]{List of observed pulsars, their observing parameters and values derived from the observations. The first three columns hold pulsar name, its period and dispersion measure, respectively. Columns 4, 5 and 6 hold the date of observation its duration and sampling time. Last two columns present flux density measurements of integrated pulse profiles and modulation indices intrinsic to the pulsar, respectively. In case where there was more than one measurement of the flux density and modulation index, we present the average value and its uncertainty.}
        \label{table1}
        \begin{tabular}{cccccccc}
          \hline
          \noalign{\smallskip}
          PSR & \it{P} & \it{DM}                 & Date of     & $t_{\mathrm{obs}}$ & $t_{\mathrm{samp}}$  & \it{S} & $m_{\mathrm{int}}$ \\
              & (s)    & ($\mathrm{cm^{-3} pc}$) & observation & (min)              & [${\mathrm{\mu s}}$] & (mJy)  &                    \\
          \noalign{\smallskip}
          \hline \hline
          \noalign{\smallskip}
          B0329+54 & 0.7145 & 26.833 & 05.12.2002 & 28  & 697.6  & 1.11$\pm$0.22 & \\
                   &        &        & 11.06.2004 & 72  & 697.6  & 2.87$\pm$0.57 & \\
          \cline{7-7}
          \noalign{\smallskip}
          \multicolumn{6}{c}{}                                   & 1.99$\pm$0.40 & 0.88$\pm$0.06 \\
          \hline
          \noalign{\smallskip}
          B0355+54 & 0.1563 & 57.142 & 26.04.2004 & 9   & 152.6  & 2.17$\pm$0.43 & -- \\
          \hline
          \noalign{\smallskip}
          B0450+55 & 0.3407 & 14.495 & 05.12.2002 & 8   & 332.6  & 0.45$\pm$0.09 & -- \\
          \hline
          \noalign{\smallskip}
          B0809+74 & 1.2922 & 6.116  & 05.12.2002 & 90  & 1261.8 & 0.64$\pm$0.13 & 1.24 \\
          \hline
          \noalign{\smallskip}
          B0823+26 & 0.5306 & 19.454 & 05.12.2002 & 20  & 518.0  & 0.87$\pm$0.17 & \\
                   &        &        & 05.12.2002 & 13  & 518.0  & 0.71$\pm$0.14 & \\
                   &        &        & 26.04.2004 & 53  & 518.0  & 0.99$\pm$0.20 & \\
          \cline{7-7}
          \noalign{\smallskip}
          \multicolumn{6}{c}{}                                   & 0.86$\pm$0.17 & 1.1$\pm$0.1 \\
          \hline
          \noalign{\smallskip}
          B0950+08 & 0.2530 & 2.958  & 26.04.2004 & 30  & 247.0  & 1.09$\pm$0.22 & 1.29 \\
          \hline
          \noalign{\smallskip}
          B1133+16 & 1.1879 & 4.864  & 05.12.2002 & 10  & 1159.8 & 0.92$\pm$0.18 & \\
                   &        &        & 05.12.2002 & 90  & 1159.8 & 0.88$\pm$0.18 & \\
                   &        &        & 26.04.2004 & 120 & 60.0   & 0.73$\pm$0.15 & \\
                   &        &        & 11.06.2004 & 47  & 60.0   & 0.79$\pm$0.16 & \\
          \cline{7-7}
          \noalign{\smallskip}
          \multicolumn{6}{c}{}                                   & 0.78$\pm$0.16 & 2.11$\pm$0.14 \\
          \hline
          \noalign{\smallskip}
          B2016+28 & 0.5579 & 14.172 & 11.06.2004 & 55  & 544.8  & 0.66$\pm$0.13 & 1.15 \\
          \hline
          \noalign{\smallskip}
          B2020+28 & 0.3434 & 24.640 & 11.06.2004 & 35  & 335.2  & 1.09$\pm$0.22 & 0.69 \\
          \hline
          \noalign{\smallskip}
          B2021+51 & 0.5291 & 22.648 & 05.12.2002 & 18  & 516.6  & 1.73$\pm$0.35 & 0.75 \\
          \hline
          \noalign{\smallskip}
          B2154+40 & 1.5252 & 70.857 & 11.06.2004 & 11  & 1489.2 & 0.13$\pm$0.03 & -- \\
          \hline
          \noalign{\smallskip}
          B2310+42 & 0.3494 & 17.276 & 11.06.2004 & 9   & 341.2  & 0.17$\pm$0.03 & -- \\
          \hline
        \end{tabular}
      \end{minipage}
    \end{table*}

  \section{Data analysis and results}

    \begin{figure}
      \includegraphics[clip,angle=-90,width=0.47\textwidth]{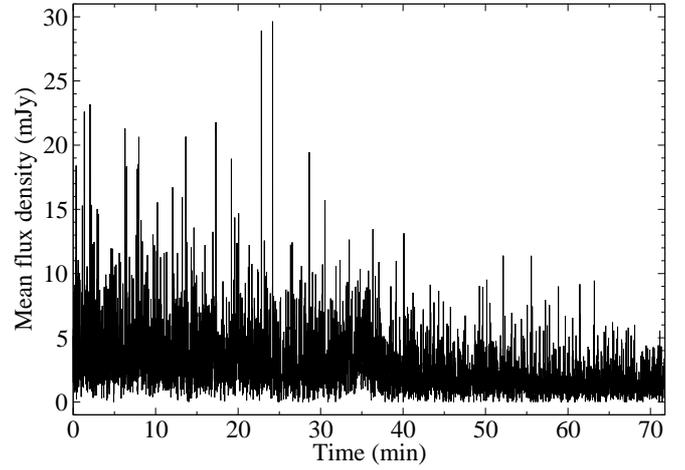}
      \caption{Calibrated time series from PSR B0329+54. Decrease of the signal strength due to the weak interstellar scintillation is clearly visible.}
      \label{0329_ts}
    \end{figure}

    \begin{figure}
      \includegraphics[clip,angle=-90,width=0.47\textwidth]{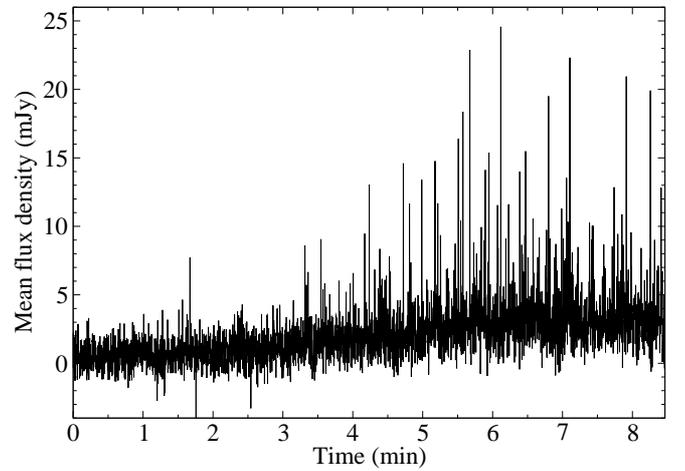}
      \caption{Calibrated time series from PSR B0355+54 showing variation of signal due to the strong interstellar scintillation causing signal to have significant variation over very short time scale.}
      \label{0355_ts}
    \end{figure}

    \begin{figure}
      \includegraphics[clip,angle=-90,width=0.47\textwidth]{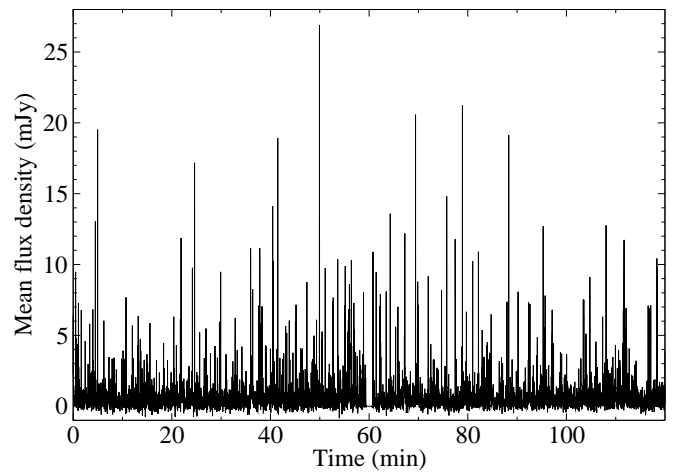}
      \caption{The longest calibrated time series (6061 pulses) from PSR B1133+16. Frequent strong single pulses can be seen. A technical break in the observations resulted in 1 minute gap in the time series visible around 60 minutes mark.}
      \label{1133_ts2}
    \end{figure}

    \subsection{Scintillation and modulation effects}

      In order to analyse the data and produce the flux-calibrated time series, we must first investigate the influence of effects which can alter the intrinsic pulsar flux. As a first effect we will consider interstellar scattering. Since, as mentioned before the pulsar scatter broadening time is strongly dependent on frequency ($\tau_{\mathrm{s}} \propto f^{-4}$, \citealt{lk05}) we expect thr effects of scattering to have little influence on our observations. The empirical relationship between $\tau_{\mathrm{s}}$ and dispersion measure, presented in the work of \citet{bcc+04}, allowed us to estimate $\tau_{\mathrm{s}}$, and confirm that this effect is negligible at high observing frequency of 8.35 GHz for pulsars used in our study.

      We now move to the second effect, interstellar scintillation. It is well known that scintillation strength changes according to observational parameters such as frequency and bandwidth at which the pulsar is observed or its dispersion measure. \citet{lk05} estimated transition frequency from strong to weak scintillation regime as a function of a pulsar's dispersion measures. They show that pulsars with the dispersion measure less than 40 $\mathrm{cm^{-3} pc}$ observed at 8.35 GHz will be in the weak scintillation regime. This is evident in Fig.~\ref{0329_ts} where a decrease of signal strength is seen through the observation. All pulsars presented in Table~\ref{table1}, except for 2 sources have met this criterion. The first pulsar, B2154+40, does not show such variations in its time series despite its large dispersion measure, 70.857 $\mathrm{cm^{-3} pc}$ probably due to fortuitous scintillation conditions during the observations. The second source, B0355+54, shows significant variations in flux density over short time scale, which can be seen in Fig.~\ref{0355_ts}. This behaviour is clearly due to the strong interstellar scintillation and makes flux measurement of this source uncertain. According to work of \citet{lkgk11}, the pulsar B0329+54 has its transition frequency between 8 and 10 GHz, which suggests that it may undergone occasional changes between weak and strong scintillation regime during our observations.

\begin{figure*}
        \centering
        \includegraphics[clip,width=0.80\textwidth]{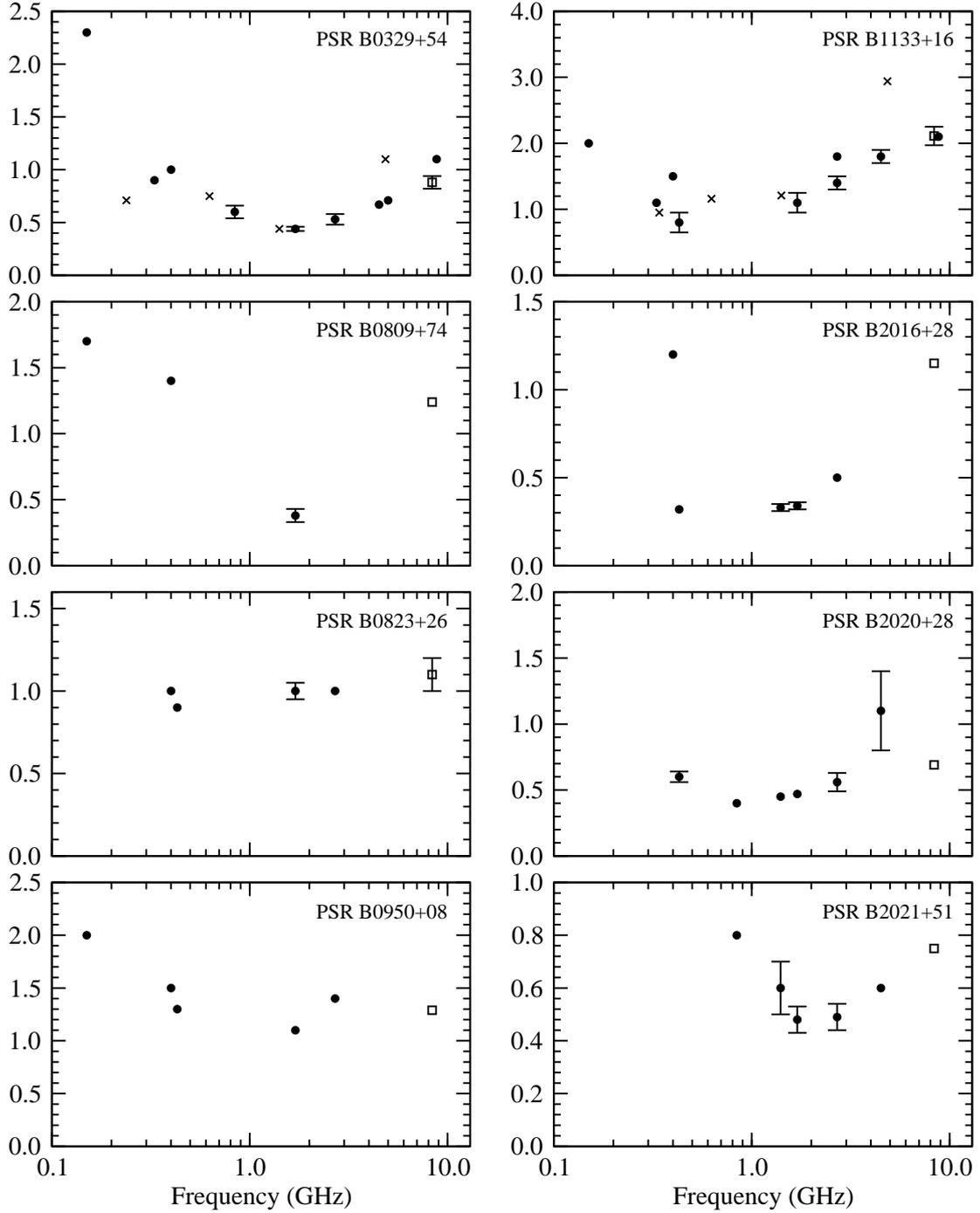}
        \caption{Modulation indices, $m$ versus frequency for the observed pulsars. It is clearly seen that $m_{\mathrm{int}}$ first decreases with frequency, and after it reaches so-called critical frequency $\nu_{\mathrm{m}}^{\mathrm{c}}$ (which for all, except B0823+26, is around 1 GHz), it rises again. Filled circles and crosses denote values taken from \citet{bsw80} and \citet{kkg+03} respectively. Open squares represent values obtained in this work.}
        \label{all_fm}
      \end{figure*}

      In order to derive the intrinsic modulation indices ($m_{\mathrm{int}}$) we used the method described by \citet{kkg+03}. Their method uses 200-s running median to correct pulsar signal for the effects of interstellar scintillation. The time scale of 200 seconds is longer than a typical pulsar intrinsic pulse-to-pulse modulation \citep[][]{wes06,wse07} and smaller than the expected scintillation times. By dividing calibrated flux of each pulse by the running median corresponding its location we corrected our observations for the effects of interstellar observations. To obtain the corrected flux density time series for interstellar scintillation effects, flux density of each pulse was divided by the running median corresponding to its location. Then the each of the data set was rescaled to be consistent with the initial average flux density. For every observation we have integrated individual pulses to obtain the average pulse profile and we calculated its flux density (see Table~\ref{table1}). In case of existing multiple observations of pulsars: B0329+54, B0823+26 and B1133+16 we also present calculated average values and their uncertainties. Our flux density measurements are consistent with values available the literature \citep[e.~g.][]{mkk+00}. Having data sets corrected for the effects of interstellar scintillation we calculated the intrinsic modulation indices, $m_{\mathrm{int}}$. We used the following equation:

      \begin{equation}
        m^{2}_{\mathrm{int}} = \frac{\left<\left(S-\left<S\right>\right)^{2}\right>}{{\left<S\right>}^{2}},
        \label{m_int}
      \end{equation}

      \noindent where {\it S} is the measurement of flux from an individual pulse, while $\left<S\right>$ is the average flux from an entire observation. Last column in Table~\ref{table1} holds calculated values of $m_{\mathrm{int}}$. In the case where a pulsar was observed multiple times, we present average $m_{\mathrm{int}}$ with its uncertainty. In case of pulsars B0355+54, B0450+55, B2154+40 and B2310+42 the observation times were too short to provide reliable estimates of $m_{\mathrm{int}}$. In Fig.~\ref{all_fm} we collate our modulation index values as a function of frequency with the results presented by \citet{bsw80} and \citet{kkg+03}. It is clearly seen that $m_{\mathrm{int}}$ first decreases with frequency, and after it reaches the so-called critical frequency $\nu_{\mathrm{m}}^{\mathrm{c}}$ (with value of $\sim$1 GHz), it rises again.

    \subsection{Null and on states}

      \begin{figure}
        \includegraphics[clip,width=0.47\textwidth]{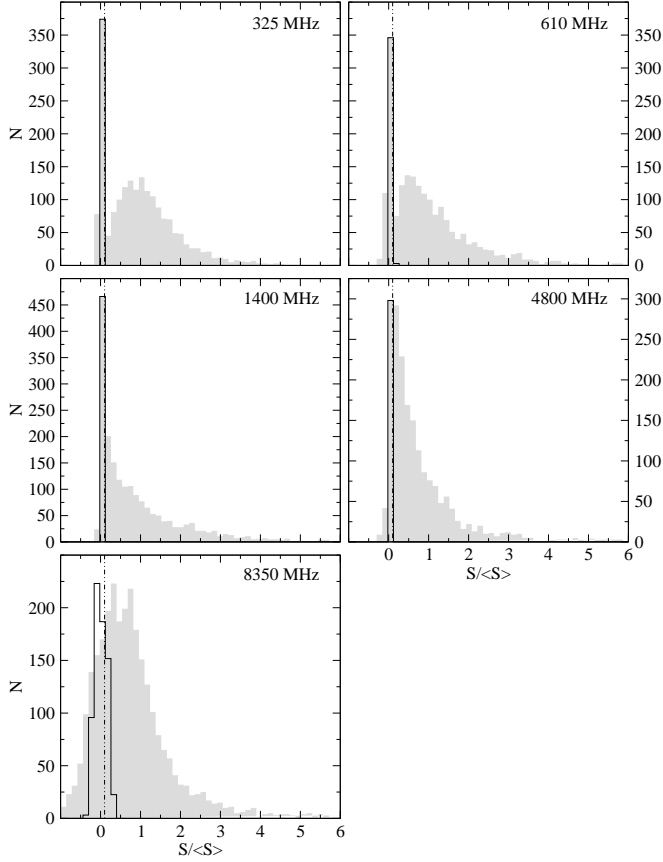}
        \caption{Pulse-energy distributions for the pulsar B1133+16. Panels presenting distribution for frequencies from 325 MHz to 4800 MHz are produced using data from \citet{bgk+07}. The panel for 8.35 GHz presents the pulse-energy distribution prepared from our data. The vertical dotted line represents an integrated-intensity threshold of ${\rm 0.10 <S>}$ to distinguish the nulls. Filled grey regions show the on-pulse energy while black contour the off-pulse energy. The offpulse noise is included for every pulse, but due to the large number of offpulse values near zero, their distributions were scaled according to the highest onpulse bin. 139 pulses with S/<S> above 6 were removed from the 8.35 GHz panel in order to keep the same scale of the plots.}
        \label{bhat}
      \end{figure}

      In Fig.~\ref{bhat} we present pulse energy distributions for PSR B1133+16 spanning over frequency range from 325 MHz to 8.35 GHz. The panels for distributions obtained from observations at lower frequencies were prepared using data from \citet{bgk+07}. The panel showing the distribution of pulse energies at 8.35 GHz is chosen from a number of similar distributions produced from our data sample. In Fig.~\ref{bhat} one can see that the distribution at lower frequencies is bimodal with classical separation into two components. At higher frequencies the distributions merge with each other into positively skewed distribution \citep{bgk+07}. The distribution around zero at lower frequencies (e.g. 325 MHz) most probably represents so called pseudo-nulls \citep{hr07}, while the right hand side of the histogram is nothing else but a Gaussian energy distribution of pulses around the pulsar energy mean value. The change of the distribution with increasing frequency is mainly caused by the decrease of average pulsar energy with the value of spectral index of -1.9 for this pulsar \citep{mkk+00}. The change of distribution character from bimodal to skewed normal is caused by increasing number of genuine nulls as a result of spectral dependency of the observed energy (decreasing with frequency) and low sensitivity of the receiver. However, careful analysis of our data sets performed by \citep{hlk12} confirms the existence of pseudo-nulls - they write "we found a clear evidence for periodic pseudonulls
in the form of apparent "null zones" in the folded modulation".
pattern.

      \begin{figure}
        \includegraphics[clip,width=0.23\textwidth]{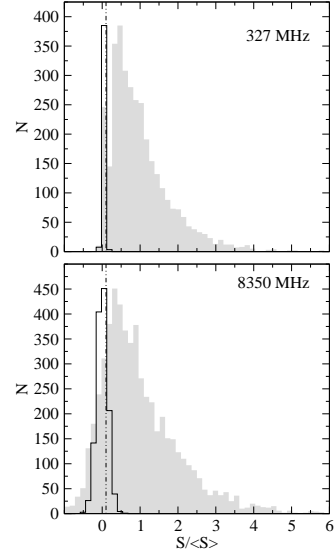}
        \caption{Pulse-energy distributions for the pulsar B0823+26. The vertical dotted line represents an integrated-intensity threshold of ${\rm 0.10 <S>}$ to distinguish the nulls. The upper panel presents distribution prepared using data from \citet{hr09} and shows the observations at 327 MHz. The lower panel shows our observations at 8.35 GHz. Filled grey regions show the on-pulse energy while black contour the off-pulse energy. The offpulse noise is included for every pulse, but due to the large number of offpulse values near zero, their distributions were scaled according to the highest onpulse bin.}
        \label{0823}
      \end{figure}
      
In Fig.~\ref{0823} we present a similar analysis of pulse-energy distributions for two peripheral observing frequencies, i.e. 327 MHz and 8.35 GHz. The panel with low frequency distribution is prepared using data used in work of \citet{hr07} while the distribution at 8.35 GHz has been produced from our data. The above mentioned effect is not so pronounced as for PSR B1133+16 which is consistent with the modulation index distribution independent of frequency. Both in Fig.~\ref{bhat} and in Fig.~\ref{0823} the raw un-normailzed data were used.

      Inspection of Fig.~\ref{bhat} (histograms of single pulse energy distributions) and its comparison with Fig.~\ref{all_fm} which shows the dependency of modulation index as a function of frequency, brings us to the conclusion that those results are related and one dependency follows the other. Namely, when we observe a "saddle" (bimodal) distribution (see histograms at 325 MHz) then the modulation index for this case has a higher value than that for 1.4 GHz where the "saddle" disappears and the distribution becomes almost symmetrical with a wide plateau around its mean value. This behaviour reflects a low value of modulation index. In case where the histogram represents a skewed distribution  (case of 8.35 GHz) the modulation index increases. Fig.~\ref{all_fm} suggests that the pseudo-nulls may be present for more pulsars than those previously mentioned (e.g. PSR B0809+74), whereas PSR B0823+26 shows the flat modulation index spectrum which may be caused by the absence of the pseudo-nulls. 
      The pulsar B0823+26 is well known for its mode changing, that is a variation between active (radio-on, hereafter) and quiescent (radio-off or null, hereafter) emission modes. In case of observations at frequencies higher thank 1 GHz, which we propose in this paper as a critical frequency, it is possible that the changes of emission modes may show the flat distribution of modulation indices as a function of the observing frequency. It is also seen in the shape of energy distribution of single pulses which may be described by a normal distribution while the measurements at lower frequencies (i.e. 327 MHz for PSR B0823+26) show the bimodal distribution. We would like to stress the importance of simultaneous observations at low and high frequencies in order to identify such behaviour. Several anomalously high points in Fig.~\ref{all_fm} may be due to nulling and mode changing phenomena as suggested by \citet{bsw80}

    \subsection{Strong single pulses in PSR B1133+16}

      In his review, \citet{cai04} defines so-called "giant pulses" as single pulses whose mean flux density exceeds average flux density by 10 $\times$ $\left<S\right>$. They also have tendency to be narrower than pulsar's average pulse, with time scales down to nano-seconds \citep[e.g.][]{hkw+03}, and occur at specific range of pulse phases, mostly at trailing edge of average pulse profile. Giant pulses are also known to show steep power-law in their energy distributions as shown by \citet{kbm+06} and \citet{kni07}.

\begin{figure}
        \includegraphics[clip,angle=-90,width=0.47\textwidth]{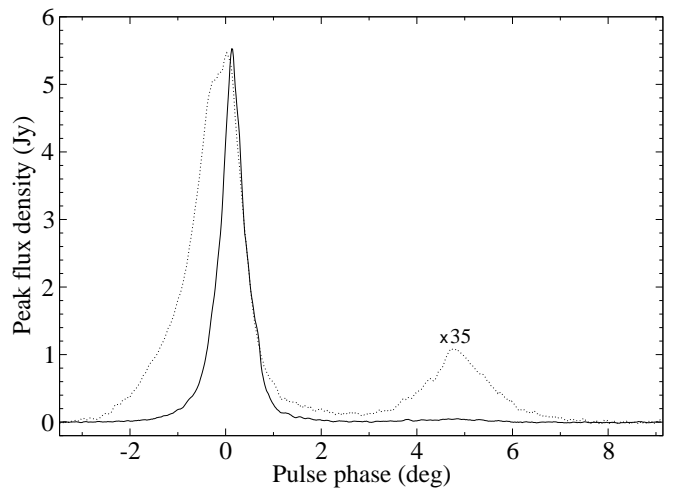}
        \caption{The average profile (solid line) made by folding strong pulses from observation of PSR B1133+16 in April 2004 relative to the integrated profile (dotted line), multiplied by a factor of 35. The tendency to occurrence of strong single pulses at the trailing edge of the first component can be seen.}
        \label{1133_avg}
      \end{figure}

\begin{figure}
        \includegraphics[clip,angle=-90,width=0.47\textwidth]{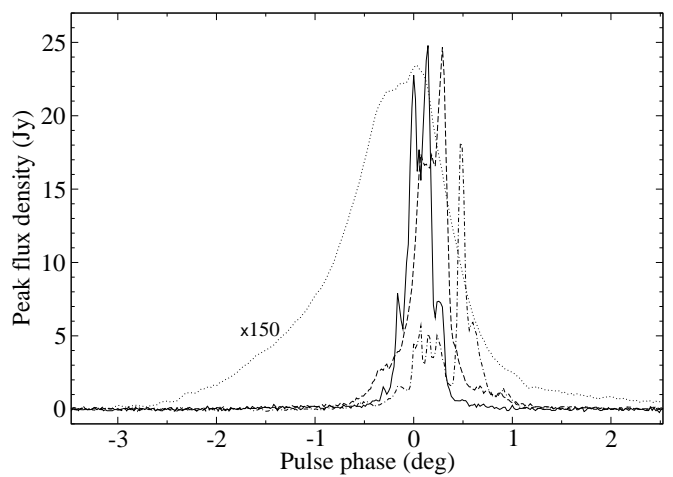}
        \caption{Three strongest pulses (solid, dashed and dot-dashed lines) from observation of PSR B1133+16 in April 2004 relative to first component of integrated profile (dotted line), multiplied by a factor of 150. Complex pulse structure and tendency to occurrence at the trailing edge of first component is shown.}
        \label{1133_3sp}
      \end{figure}

\begin{figure}
        \includegraphics[clip,angle=-90,width=0.47\textwidth]{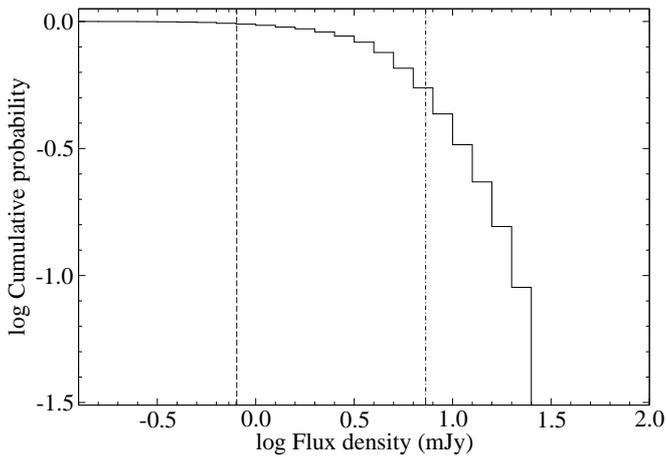}
        \caption{Cumulative probability function of single pulse flux density of PSR B1133+16 for data presented in Fig.~\ref{1133_ts2}. The dashed line denotes the mean value of flux density, $\left<S\right>$, while the dot-dashed line indicates $10 \times \left<S\right>$.}
        \label{1133_cpf}
      \end{figure}

      Pulsar B1133+16 is well known for emitting strong single pulses visible at wide range of frequencies \citep[e.g.][]{kss11, kkg+03}. During our analysis we also detected strong single pulses originating from PSR B1133+16. The strongest detected pulse was about 48 $\times$ $\left<S\right>$. Fig.~\ref{1133_ts2} presents the calibrated time series of B1133+16 with clearly visible frequently occurring strong pulses. In all observational sessions we detected 156 pulses originating from this pulsar, whose mean flux density exceeds 10 $\times$ $\left<S\right>$. During the longest observations of PSR B1133+16 done in April 2004 a total of 54 strong single pulses were detected. Fig.~\ref{1133_avg} shows the integrated pulse profile (dotted line) magnified by a factor of 35 and the pulse profile resulting from averaging only the strong single pulses observed in April 2004. One can easily see that the peak of this profile is located at the trailing edge of first component of the average profile. Fig.~\ref{1133_3sp} shows the three strongest observed single pulses from our longest observation. It is easily seen that they present very complicated structure with multiple narrow components. Despite the fact that these pulses meet "working definition" of giant pulses we can not acknowledge them as "classical" giant pulses produced in the outer gap region. In Fig.~\ref{1133_cpf} we show computed cumulative probability distribution of single pulse flux densities for the longest observation of PSR B1133+16 done in April 2004. In this plot we indicate the mean flux density $\left<S\right>$ by a dashed line and a dotted-dashed line marks the value of 10 $\times$ $\left<S\right>$. There is no noticeable evidence for any change in a slope of cumulative probability function and hence, power-law in their energy distribution. The lack of typical steep power law in the distribution prevents us from classifying them as "classical" giant pulses.

      We also report the detection of strong single pulses ($S > 10 \times \left<S\right>$) from four other pulsars during our observations: two pulses for PSR B0329+54, five pulses for PSR B0355+54, two pulses for PSR B0823+26 and three pulses for PSR B0950+08. 

  \section{Discussion and conclusions}

    The research presented in this paper concerned pulse-to-pulse flux density modulation of pulsars at high frequency. We have measured flux densities of integrated pulse profiles for 12 pulsars, which agreed with the known literature values \citep{mkk+00}. For multiple observational sessions available for three pulsars, we have calculated average values of flux densities and their uncertainties. For two pulsars we have encountered long term variations of their flux density time series. PSR B0355+54 has shown very strong variations over very short time scales (i.e. 9 minutes). This resulted in large uncertainty in flux density (see Table~\ref{table1}). The origin of the variations can be assigned to the strong interstellar scintillations. 

    Analysis of behaviour of modulation indices as a function frequency has shown that its value in most cases has its minimum around 1 GHz. The distribution changes from bimodal to normal symmetrical with its maximum around zero. The changes of pulse-energy distributions of the observed pulses result in changes of modulation indices depending on frequency. Bimodal distribution at lower frequencies occurs because the maximum around zero represents so called pseudo-nulls (no detection due to properties of the radiation beam structure) and the right side of the histogram is nothing else but a Gaussian pulse-energy distribution around the average value. The change of such bimodal distribution into the normal one with maximum around zero with increasing observing frequency is mainly caused by decreasing average pulsar energy with the approximate value of spectral index of -2.0. Therefore the number of so called genuine nulls (no detection due to inadequate receiver sensitivity) increases.

    We report the detection of strong single pulses from PSR B1133+16. Although they meet "working definition" of giant pulses, (i.e. 10 $\times$ $\left<S\right>$ and constrains on appearance at trailing edge of first component), complexity of structure in single pulses and no visible power-law behaviour in their distribution energy suggest other phenomena responsible for such behaviour. It is possible that those "giant pulses" come from polar cap as suggested by \citep{gm05}. Therefore longer observation with grater resolution are required to obtain better signal-to-noise ratio and more detailed structure of single pulses for this pulsar. This will show if there exists a power-law in energy distribution and give us a clue for existence of giant-pulse phenomenon in this source.

  \begin{acknowledgements}
    This paper is based on observations with the 100-m telescope of the MPIfR (Max-Planck-Institut f\"ur Radioastronomie) at Effelsberg. Authors would like to thank Michael Kramer for valuable suggestions during data analysis and W. Lewandowski for fruitful discussions. MS would like to thank Joanna Rankin, Jeffrey Herfindal and Ramesh Bhat for providing the data used to calculate pulse-energy distributions. This paper was supported by the grant DEC-2012/05/B/ST9/03924 of the Polish National Science Centre. This work made use of the NASA ADS astronomical data system.
  \end{acknowledgements}

  \bibliographystyle{aa}
  \bibliography{pulse_ms_12.10-rev}

\end{document}